\documentclass[letterpaper,journal]{IEEEtran}
\usepackage{amsmath,amssymb,amsfonts}
\usepackage{array}
\usepackage[caption=false,font=normalsize,labelfont=sf,textfont=sf]{subfig}
\usepackage{textcomp}
\usepackage{stfloats}
\usepackage{url}
\usepackage{verbatim}
\usepackage{graphicx}
\usepackage{cite}
\usepackage{bm}

\usepackage{xcolor}
\usepackage{balance}
\usepackage{amsthm}

\newtheorem{remark}{Remark}

\begin{document}
\raggedbottom
\setlength{\textfloatsep}{5pt plus 1pt minus 2pt}
\setlength{\dbltextfloatsep}{5pt plus 1pt minus 2pt}
\setlength{\floatsep}{5pt plus 1pt minus 2pt}
\setlength{\intextsep}{5pt plus 1pt minus 2pt}
\setlength{\abovecaptionskip}{1pt}
\setlength{\belowcaptionskip}{-2pt}

\title{Extended-Target Classification and Localization \\for Near-Field ISAC}

\author{Zongyao Zhao, Zhaolin Wang, Lincong Han, Jing Jin, \\ Yuanwei Liu,~\IEEEmembership{Fellow, IEEE}, and Kaibin Huang,~\IEEEmembership{Fellow, IEEE}
\thanks{An early version of this work is available on arXiv \cite{ZhaoICCW2026} and has been accepted for presentation at an IEEE International Conference on Communications (ICC) 2026 Workshop. \emph{(Corresponding authors: Yuanwei Liu and Kaibin Huang)}}

\thanks{Z. Zhao, Z. Wang, Y. Liu, and K. Huang are with the Department of Electrical and Computer Engineering, The University of Hong Kong, Hong Kong (E-mail: zongyao@hku.hk, zhaolin.wang@hku.hk, yuanwei@hku.hk, huangkb@hku.hk).}

\thanks{L. Han and J. Jin are with Future Research Laboratory, China Mobile Research Institute, Beijing (E-mails: hanlincong@chinamobile.com, jinjing@chinamobile.com).}

}
\markboth{}%
{Zhao \MakeLowercase{\textit{et al.}}: Extended-Target Classification and Localization for Near-Field ISAC}

\maketitle
\thispagestyle{empty}

\begin{abstract}
Near-field integrated sensing and communication (ISAC) enables object-level sensing from distance-dependent array responses, yet most existing near-field methods still rely on point-target models and realistic extended targets remain largely unexplored. In this paper, joint target classification and range-azimuth localization are studied from channel responses of realistic extended targets. A dual-branch inference framework is proposed. Semantic and geometric branches are used for classification and localization, respectively. Cross-task attention is introduced after task-specific encoding so that complementary cues can be exchanged without forcing full feature sharing from the input stage. To improve localization on the same backbone, uncertainty-aware regression and a physics-guided structured objective are adopted, including planar consistency, peak-response regularization, and geometry-coupling constraints. Training and evaluation data are generated from full-wave electromagnetic scattering simulations of voxelized vehicle targets with randomized heading angles, material contrasts, and placements. The compared variants show that cross-task attention mainly benefits classification, while uncertainty-aware and structured supervision are needed to recover strong localization performance on the same backbone. Under the adopted shared-OFDM benchmark, the proposed framework reaches the best joint operating point with fewer sensing tones for the same target performance region.
\end{abstract}


\section{Introduction}
\IEEEPARstart{I}ntegrated sensing and communication (ISAC) is a key capability for sixth-generation (6G) systems, where sensing and communication share hardware, spectrum, and signal processing resources \cite{Hassan2016,Chiri2017,Hassan2019,Cui2021,Liu2022,Liu2019,Zhang2021JSTSP}. A central challenge in ISAC is the tradeoff between communication efficiency and sensing performance under a shared time-frequency budget. This tradeoff becomes particularly acute for object-level sensing, which demands richer observations and hence places heavier pressure on limited sensing resources.

To manage this tradeoff, existing ISAC studies have mainly focused on waveform and beamforming design, signal processing, and communication--sensing tradeoff optimization. Transmit covariance matrices, beam patterns, and probing sequences have been designed to support sensing and communication simultaneously \cite{LiuX2020,Cheng2021,Zhao2022}. Uncertainty-aware beamforming formulations have also been explored for ISAC transceiver design \cite{ZhaoBayesian2025}. Cram\'er-Rao-bound-based beamforming and design methods have been developed for sensing-aware transceiver optimization \cite{LiuF2022,ZhaoCRB2024}. Joint optimization of communication rate and sensing quality has also been studied under shared resource constraints in several ISAC settings, including bistatic, multi-user, and multi-target scenarios \cite{Guo2023,Zhu2023,ZhaoMultiTarget2025}. Most of them, however, still adopt far-field propagation and point-target models. The sensing outputs are mainly geometric parameters such as range, angle, and velocity rather than object-level semantic information. As a result, the target is usually represented by a few low-dimensional parameters instead of a structured object with category-dependent scattering behavior.

As the array aperture grows from conventional massive multiple-input multiple-output (MIMO) to extremely large-scale antenna systems, many targets enter the radiating near-field region. In this regime, spherical-wave propagation replaces the far-field plane-wave model and yields distance-dependent array signatures \cite{Wang2023,Wang2025NFLS}. The resulting angle-range coupling enables higher-resolution sensing and has stimulated extensive research on near-field localization and sensing \cite{Wang2025NFLS,Chen2024NFSurvey}. Recent near-field ISAC studies have developed signal models and sensing frameworks with distance-dependent responses \cite{Wang2023,Zhao2024NFISAC}, integrated sensing-positioning-communication formulations \cite{Li2024ISPAC}, performance analysis and beam-focusing designs \cite{Qu2024NFISAC}, and system-level transceiver or beamforming designs for near-field extremely large-scale multiple-input multiple-output (XL-MIMO) systems \cite{SunNOMA2025,Meng2025Modular}. Related near-field XL-MIMO studies have also considered joint tasks such as activity detection, channel estimation, and localization \cite{Qiao2024XL}. These results confirm that near-field array responses carry richer geometric information than their far-field counterparts. Despite this progress, most existing works still address point targets and geometric estimation, where the target response can still be described by a small number of physical parameters.

\begin{figure}[!t]
\centering
\includegraphics[width=0.48\textwidth]{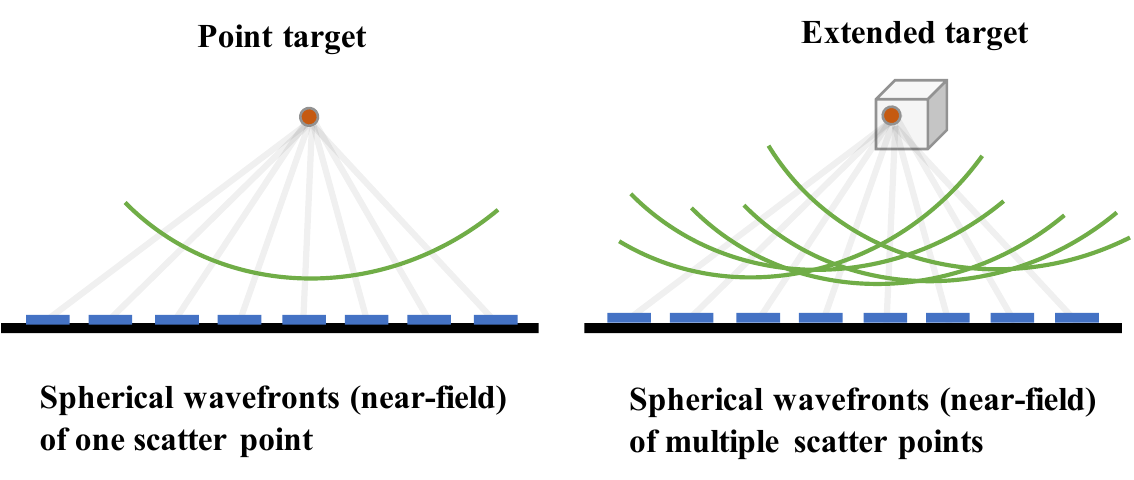}
\caption{Illustration of near-field point-target and extended-target sensing.}
\label{fig:point_extended_nearfield}
\end{figure}

Practical objects are extended targets rather than isolated point reflectors in near-field scenarios. Their echoes are produced by many distributed scatterers whose amplitudes and phases depend on target shape, material composition, and pose. The received near-field channel is a coupled superposition of many scattering contributions across the array aperture and the sensing subcarriers. For a point target, a low-dimensional spherical-wave model is often sufficient, and the resulting angle-range coupling can be exploited directly for model-based localization \cite{Bekkerman2006,Kay1993}. For realistic extended targets, such a tractable model is usually unavailable. Applying point-target methods then leads to severe model mismatch. Point-target sensing follows a simple spherical-wave response; extended-target sensing is shaped by multiple interacting scatterers (Fig.~\ref{fig:point_extended_nearfield}). This motivates direct inference from near-field ISAC channel responses. Existing near-field ISAC studies with extended targets are still limited. Extremely-large-aperture-array (ELAA)-based near-field ISAC performance in the presence of extended targets and clutter has been analyzed in \cite{Dassanayake2025ELAA}, while \cite{Hu2025Beamfocusing} studies energy-efficient hybrid beamfocusing under both point-target and extended-target formulations. Electromagnetic modeling results in \cite{Sambon2025EM} further analyze when point-target approximations remain adequate and when model mismatch emerges as target extent and shape are explicitly accounted for. These studies mainly address performance analysis, electromagnetic modeling, or beam design rather than receiver-side joint semantic recognition and geometric localization, where the two tasks rely on different cues and are not well handled by naive feature sharing.

Classification-only recognition is a useful first step, and a preliminary version of this research studied classification-only near-field target recognition \cite{ZhaoICCW2026}. Practical sensing, however, also requires target location for downstream functions such as scene understanding, tracking initialization, and sensing-assisted communication control. This paper extends that early result to joint classification and range-azimuth localization. The key difficulty is that the two tasks do not rely on the same cues. Classification depends more on morphology and material related signatures, while localization depends more on phase and spatial-frequency variations induced by target position. Joint inference is therefore more challenging than solving either task alone and is not well handled by naive feature sharing.

This paper investigates receiver-side joint target classification and localization from near-field ISAC channel responses of realistic extended targets under limited frequency resources. The first goal is joint semantic-geometric inference beyond point-target abstractions, while the second is to determine how much sensing bandwidth is required to reach a target sensing-quality region under a shared orthogonal frequency-division multiplexing (OFDM) budget. The design starts from a simple observation: category and position are embedded in the same channel tensor, but they are not expressed through the same structures. Category is tied more strongly to distributed scattering patterns, while location is tied more strongly to phase evolution and array-frequency geometry. The proposed network therefore does not enforce hard sharing from the first layer. Instead, two task-specific branches are used first, and bidirectional cross-task attention is introduced only after branch-wise encoding. Geometric cues can then assist semantic recognition without fully collapsing the two representations. To recover localization quality on the same CTA backbone, uncertainty-aware regression and a physics-guided structured objective are further introduced for range-azimuth estimation. A benchmark-rate calibration is included to interpret the bandwidth implication under the shared sensing-communication budget.

The main contributions of this paper are summarized as follows:
\begin{itemize}
    \item A physics-grounded joint inference problem is formulated for realistic near-field extended targets, where category and range-azimuth location are inferred from limited multi-frequency observations. The formulation makes explicit that semantic recognition and geometric localization depend on different structures in the same channel tensor and are therefore not well served by naive feature sharing.
    \item A task-aware dual-branch framework is developed. Semantic and geometric branches first preserve task-specific statistics, and token-level cross-task attention is introduced only after branch-wise encoding to moderate task interference while still enabling selective information exchange.
    \item An uncertainty-aware and physics-guided objective is developed as structured inductive bias for the same CTA backbone. The experiments show that CTA mainly improves classification, whereas the additional structured terms are needed to recover localization on the shared backbone. Under the adopted shared-OFDM rate interpretation, the resulting model reaches the same sensing-QoS region with fewer sensing tones.
\end{itemize}

{\em Notation}: Italic letters denote scalars; bold lowercase and uppercase letters denote vectors and matrices, respectively; and calligraphic letters denote sets or higher-order tensors. For a vector or matrix, $[\cdot]_i$ and $[\cdot]_{m,n}$ denote its entries, and $|\cdot|$ denotes either the absolute value of a scalar or the cardinality/measure of a set according to context. $\mathbb{R}$ and $\mathbb{C}$ denote the real and complex fields. $(\cdot)^{\mathsf T}$ and $(\cdot)^{\mathsf H}$ denote the transpose and conjugate transpose, while $\Re\{\cdot\}$ and $\Im\{\cdot\}$ denote the real and imaginary parts. $\mathbb{E}[\cdot]$ denotes expectation, $\mathcal{CN}(\cdot,\cdot)$ denotes the circularly symmetric complex Gaussian distribution, and $\bm I_n$ denotes the $n\times n$ identity matrix. $\|\cdot\|_1$ and $\|\cdot\|_2$ denote the $\ell_1$ and Euclidean norms, respectively, and $\triangleq$ denotes equality by definition. A hat indicates an estimate or network prediction, while a tilde is used for transformed or real-valued representations such as the channel tensor $\tilde{\mathcal H}_s$. Unless otherwise specified, subscripts $t$, $r$, $k$, and $i$ index transmit antennas, receive antennas, subcarriers, and samples, respectively, and all angles are measured in radians.

\section{System Model and Problem Formulation}\label{sec2}

\begin{figure}[!t]
\centering
\includegraphics[width=\columnwidth]{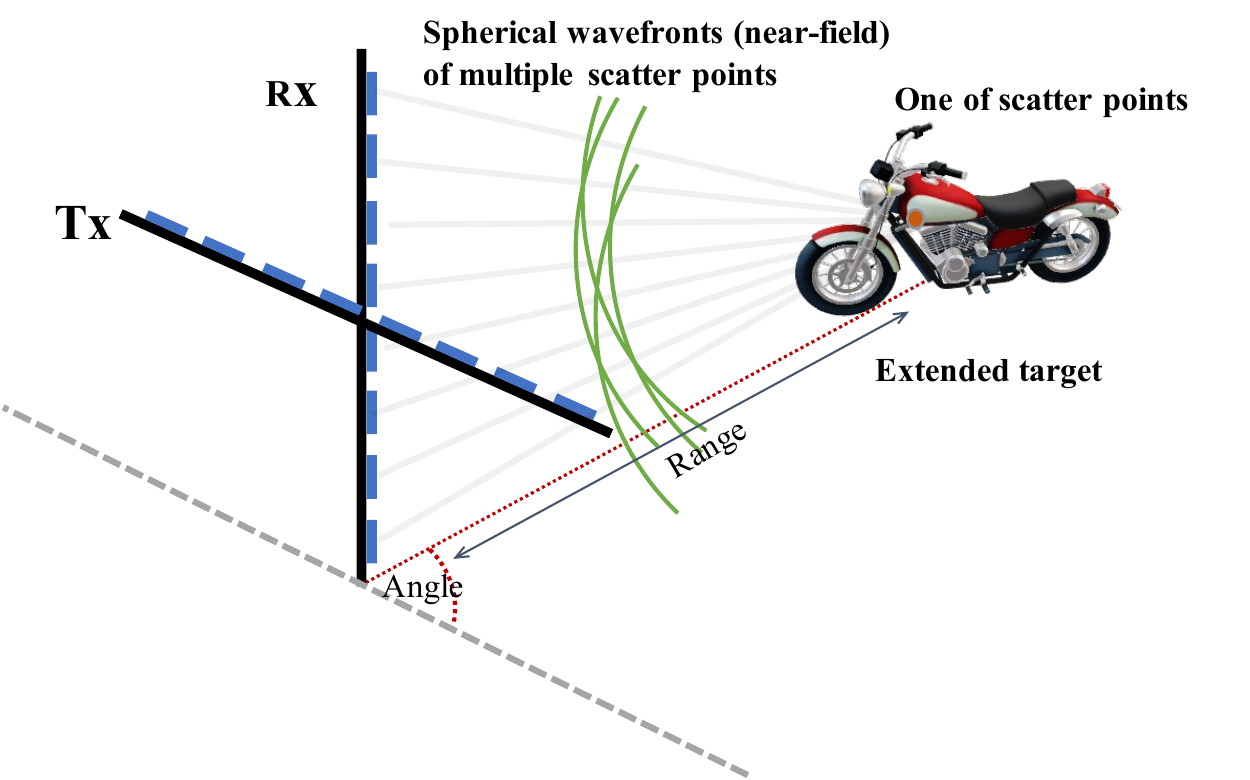}
\caption{Illustration of the near-field extended-target sensing scenario.}
\label{fig:extended_nearfield}
\end{figure}

We consider a near-field integrated sensing and communication (ISAC) base station equipped with a cross-shaped multiple-input multiple-output (MIMO) array comprising $N_t$ transmit and $N_r$ receive antennas. The transmit uniform linear array (ULA) is deployed along the $y$-axis and the receive ULA along the $z$-axis, both with element spacing $d=\lambda_c/2$, where $\lambda_c$ is the carrier wavelength. Owing to the large array aperture, the target of interest is assumed to lie in the radiating near-field region \cite{Wang2023,Wang2025NFLS}. Let $\bm r_t^{(\mathrm{tx})}\in\mathbb{R}^3$ and $\bm r_r^{(\mathrm{rx})}\in\mathbb{R}^3$ denote the positions of the $t$-th transmit and $r$-th receive antennas, respectively; see Fig.~\ref{fig:extended_nearfield}.
An orthogonal frequency-division multiplexing (OFDM) waveform with $K$ subcarriers is employed \cite{Li2024ISPAC,Zhao2024NFISAC}. On the $k$-th subcarrier at frequency $f_k$ and angular frequency $\omega_k=2\pi f_k$, the probing vector is denoted by $\bm x_k\in\mathbb{C}^{N_t\times 1}$. The corresponding received echo is modeled as
\begin{align}
    \bm y_k = \bm H_k\,\bm x_k + \bm z_k,
    \label{eq:sensing_signal}
\end{align}
where $\bm H_k\in\mathbb{C}^{N_r\times N_t}$ is the target-induced near-field MIMO channel matrix and $\bm z_k\sim\mathcal{CN}(\bm 0,\sigma_s^2\bm I_{N_r})$ denotes additive noise. The formulation focuses on the residual target response after background calibration or suppression, so dominant environmental echoes are assumed removed in preprocessing. In this work, $\{\bm x_k\}$ are assumed known, and the estimated channel responses $\{\bm H_k\}$ are directly used as input features for joint target classification and localization.

The scattering model is three-dimensional, whereas the localization task is restricted to range-azimuth estimation in a prescribed sensing plane. Let $\Pi:\mathbb{R}^3\rightarrow\mathbb{R}^2$ denote the projection from the three-dimensional global coordinate to the local coordinate in this plane, and write $\bar{\bm r}=\Pi(\bm r)$ for the projected coordinate of point $\bm r$.

\subsection{Near-Field Extended-Target Scattering Model}

Unlike idealized point-target models, realistic targets occupy a finite spatial region and generate distributed scattering. Let $V_s\subset\mathbb{R}^3$ denote the target support region. The target is modeled as an isotropic inhomogeneous dielectric scatterer with spatially varying relative permittivity $\varepsilon_r(\bm r)$ and conductivity $\sigma(\bm r)$. Under standard frequency-domain dielectric-scattering formulations \cite{Martin1998}, the complex contrast on subcarrier $k$ is defined as
\begin{align}
    \chi_k(\bm r)
    =
    \varepsilon_r(\bm r)-1
    -
    j\frac{\sigma(\bm r)}{\omega_k\varepsilon_0},
    \qquad \bm r\in V_s,
    \label{eq:contrast_final}
\end{align}
where $\varepsilon_0$ is the free-space permittivity. The background medium outside $V_s$ is assumed to be homogeneous free space with permittivity $\varepsilon_0$ and permeability $\mu_0$.

\subsubsection{Dipole Excitation and Incident Field}

Each transmit element is modeled as a small electric dipole with effective dipole moment $\bm p_t\in\mathbb{C}^3$ located at $\bm r_t^{(\mathrm{tx})}$. For a dipole source at $\bm r_{\mathrm{src}}$ with moment $\bm p$ and an observation point $\bm r$, define
\begin{align}
    R=\|\bm r-\bm r_{\mathrm{src}}\|,\qquad
    \widehat{\bm R}=\frac{\bm r-\bm r_{\mathrm{src}}}{R}.
\end{align}
At subcarrier $k$, the free-space wavenumber is $k_0=\omega_k\sqrt{\mu_0\varepsilon_0}$. The electric field generated by the dipole in free space is written as
\begin{align}
    \bm E_k^{\mathrm{dip}}(\bm r;\bm r_{\mathrm{src}},\bm p)
    &=
    \frac{e^{-jk_0R}}{4\pi\varepsilon_0}
    \left[
    \frac{k_0^2}{R}\left(\bm I_3-\widehat{\bm R}\widehat{\bm R}^{\mathsf T}\right)
    \right.\nonumber\\
    &\quad\left.
    +
    \left(\frac{1}{R^3}-\frac{jk_0}{R^2}\right)
    \left(3\widehat{\bm R}\widehat{\bm R}^{\mathsf T}-\bm I_3\right)
    \right]\bm p,
    \label{eq:dipole_field_final}
\end{align}
where $\bm I_3$ is the $3\times 3$ identity matrix.

Stacking the incident fields generated by all transmit antennas, we define the incident-field transfer matrix at position $\bm r$ as
\begin{align}
    \bm A_{k,\mathrm{inc}}(\bm r)
    &=
    \big[
    \bm E_k^{\mathrm{dip}}(\bm r;\bm r_1^{(\mathrm{tx})},\bm p_1),\,
    \ldots,\,\nonumber\\
    &\qquad
    \bm E_k^{\mathrm{dip}}(\bm r;\bm r_{N_t}^{(\mathrm{tx})},\bm p_{N_t})
    \big]
    \in\mathbb{C}^{3\times N_t}.
    \label{eq:Ainc_final}
\end{align}
For a given probing vector $\bm x_k$, the superposed incident field is
\begin{align}
    \bm E_{k,\mathrm{inc}}(\bm r)=\bm A_{k,\mathrm{inc}}(\bm r)\bm x_k.
    \label{eq:Einc_final}
\end{align}

\subsubsection{Volume Integral Equation and Total-Field Transfer}

Under standard assumptions for inhomogeneous dielectric scattering, the total electric field inside $V_s$ satisfies the volume integral equation (VIE) \cite{Martin1998}
\begin{align}
    \bm E_k(\bm r)
    &=
    \bm E_{k,\mathrm{inc}}(\bm r)
    +
    k_0^2\int_{V_s}
    \overline{\overline{\bm G}}_k(\bm r,\bm r')\nonumber\\
    &\quad\times
    \chi_k(\bm r')\bm E_k(\bm r')
    \,{\rm d}\bm r',
    \qquad \bm r\in V_s,
    \label{eq:vie_final}
\end{align}
where $\overline{\overline{\bm G}}_k(\bm r,\bm r')$ is the free-space dyadic Green's function \cite{Arnoldus2001}
\begin{align}
    \overline{\overline{\bm G}}_k(\bm r,\bm r')
    =
    \left(
    \bm I_3+\frac{1}{k_0^2}\nabla\nabla^{\mathsf T}
    \right)
    g_k(\|\bm r-\bm r'\|),
    \label{eq:dyadic_green_final}
\end{align}
with scalar Green's function
\begin{align}
    g_k(R)=\frac{e^{-jk_0R}}{4\pi R}.
    \label{eq:scalar_green_final}
\end{align}

Since \eqref{eq:vie_final} is linear in $\bm x_k$, the total field depends linearly on the probing vector. Therefore, for each $\bm r\in V_s$, there exists a total-field transfer matrix $\bm A_k(\bm r)\in\mathbb{C}^{3\times N_t}$ such that
\begin{align}
    \bm E_k(\bm r)=\bm A_k(\bm r)\bm x_k.
    \label{eq:Adef_final}
\end{align}
Substituting \eqref{eq:Einc_final} and \eqref{eq:Adef_final} into \eqref{eq:vie_final} yields
\begin{align}
    \bm A_k(\bm r)
    &=
    \bm A_{k,\mathrm{inc}}(\bm r)
    +
    k_0^2\int_{V_s}
    \overline{\overline{\bm G}}_k(\bm r,\bm r')\nonumber\\
    &\quad\times
    \chi_k(\bm r')\bm A_k(\bm r')
    \,{\rm d}\bm r',
    \qquad \bm r\in V_s.
    \label{eq:A_vie_final}
\end{align}
\subsubsection{Receiver Model and Channel Matrix Construction}

Let $\bm q_r\in\mathbb{C}^3$ denote the receive polarization vector of the $r$-th antenna. Stacking all receive elements, define
\begin{align}
    \bm B_k(\bm r')
    =
    \begin{bmatrix}
        \bm q_1^{\mathsf H}\overline{\overline{\bm G}}_k(\bm r_1^{(\mathrm{rx})},\bm r')\\
        \vdots\\
        \bm q_{N_r}^{\mathsf H}\overline{\overline{\bm G}}_k(\bm r_{N_r}^{(\mathrm{rx})},\bm r')
    \end{bmatrix}
    \in\mathbb{C}^{N_r\times 3}.
    \label{eq:B_final}
\end{align}
For notational simplicity, constant antenna and hardware scaling factors are absorbed into the definition of the channel matrix. Then the target-induced channel response on subcarrier $k$ is written as
\begin{align}
    \bm H_k
    =
    k_0^2\int_{V_s}
    \bm B_k(\bm r')
    \chi_k(\bm r')
    \bm A_k(\bm r')
    \,{\rm d}\bm r'.
    \label{eq:Hk_cont_final}
\end{align}
The observed channel response is determined by distributed scattering over the target support rather than by a single ideal reflection point.

For numerical implementation, $V_s$ is discretized into $N_s$ voxels with centers $\{\bm r_n\}_{n=1}^{N_s}$ and volume $\Delta V$. Then \eqref{eq:Hk_cont_final} is approximated as
\begin{align}
    \bm H_k
    \approx
    k_0^2\Delta V
    \sum_{n=1}^{N_s}
    \bm B_k(\bm r_n)\chi_k(\bm r_n)\bm A_k(\bm r_n).
    \label{eq:Hk_disc_final}
\end{align}
The matrices $\{\bm H_k\}$ constitute the raw near-field ISAC observations and are generated offline by solving the discretized forward scattering problem.

\subsection{Channel Tensor Representation}

Let $\mathcal K_s=\{k_1,\ldots,k_{K_s}\}$ denote the selected subcarrier set, where $1\le K_s\le K$ and $k_1<\cdots<k_{K_s}$. The corresponding multi-frequency observation is stacked as
\begin{align}
    \mathcal H_s
    =
    \big[
    \bm H_{k_1},\bm H_{k_2},\ldots,\bm H_{k_{K_s}}
    \big]
    \in
    \mathbb{C}^{N_r\times N_t\times K_s}.
    \label{eq:Hstack_final}
\end{align}
For neural processing, the real and imaginary parts are stacked to form the real-valued tensor
\begin{align}
    \tilde{\mathcal H}_s
    \in
    \mathbb{R}^{2\times N_r\times N_t\times K_s},
    \label{eq:Hreal_final}
\end{align}
with
\begin{align}
    [\tilde{\mathcal H}_s]_{1,:,:,:}=\Re\{\mathcal H_s\},\qquad
    [\tilde{\mathcal H}_s]_{2,:,:,:}=\Im\{\mathcal H_s\},
\end{align}
where the first dimension corresponds to the real and imaginary channels. This representation keeps the receive, transmit, and frequency dimensions explicit for subsequent inference.

\subsection{Communication Model and Resource Partition}

To interpret the sensing allocation from the communication perspective, we embed the selected sensing tones into a shared OFDM block with $K_{\mathrm{tot}}$ total subcarriers. Among them, $K_s$ tones are reserved for sensing and the remaining
\begin{align}
    K_c = K_{\mathrm{tot}} - K_s
\end{align}
tones are available for communication. Here, $K_{\mathrm{tot}}$ is a system-level OFDM size introduced to calibrate the sensing-communication tradeoff and does not alter the sensing tensor dimension used by the learning model.
After standard communication precoding and combining, let $h_k$ denote the effective scalar downlink channel coefficient on active communication tone $k$. We assume independent and identically distributed (i.i.d.) Rayleigh fading, i.e., $h_k\sim\mathcal{CN}(0,1)$, and equal power allocation across the $K_c$ active communication tones. Let $\bar\gamma_c$ denote the average received signal-to-noise ratio (SNR) per communication subcarrier when the full OFDM block is used for communication. Once $K_s$ tones are reserved for sensing, the redistributed power over the remaining communication tones yields the per-tone instantaneous SNR $\bar\gamma_c K_{\mathrm{tot}}|h_k|^2/K_c$. We define the normalized ergodic-rate benchmark as
\begin{align}
    R_{\mathrm{erg}}(K_s)
    &\triangleq
    \mathbb{E}\!\left[
    \frac{1}{K_{\mathrm{tot}}}
    \sum_{k=1}^{K_c}
    \log_2\!\left(
    1+
    \bar\gamma_c\frac{K_{\mathrm{tot}}}{K_c}|h_k|^2
    \right)
    \right],
    \label{eq:Rerg}
\end{align}
with units of bit/s/Hz over the entire OFDM block. It captures both the loss of communication tones and the compensating power concentration on the remaining tones.

This leads to a simple sensing-bandwidth objective. Let $A(K_s)$ denote the achievable classification accuracy and let $E_{\mathrm p}(K_s)$ denote the achievable planar localization error when $K_s$ sensing tones are reserved. Since larger $K_s$ leaves fewer tones for communication under \eqref{eq:Rerg}, maximizing the residual communication rate is equivalent to minimizing the sensing bandwidth subject to sensing-QoS constraints:
\begin{align}
    K_s^\star
    =
    \arg\max_{K_s\in\{1,\ldots,K\}} &\, R_{\mathrm{erg}}(K_s)
    =
    \arg\min_{K_s\in\{1,\ldots,K\}} K_s
    \notag\\
    \text{s.t.}\quad
    &A(K_s)\ge \tau_{\mathrm{cls}},\ 
    E_{\mathrm p}(K_s)\le \tau_{\mathrm{loc}}.
    \label{eq:qos_rate_alloc}
\end{align}
\eqref{eq:qos_rate_alloc} can be used to compare which inference architecture reaches a desired sensing-quality region with the smallest sensing bandwidth under the shared OFDM budget. Accordingly, $R_{\mathrm{erg}}(K_s)$ provides a communication-aware interpretation of bandwidth-efficient sensing.

\subsection{Joint Classification and Localization Task}

Let $c\in\{1,\ldots,C\}$ denote the target class label. The localization label is defined by the geometric center of the target support region,
\begin{align}
    \bm r_{\mathrm{ctr}}
    =
    \frac{1}{|V_s|}
    \int_{V_s}\bm r\,{\rm d}\bm r
    \approx
    \frac{1}{N_s}\sum_{n=1}^{N_s}\bm r_n,
    \label{eq:center_label}
\end{align}
where $|V_s|$ denotes the target volume. Let $\bar{\bm r}_{\mathrm{ctr}}=[x_c,y_c]^{\mathsf T}\in\mathbb{R}^2$ denote the projected center coordinate, i.e., $\bar{\bm r}_{\mathrm{ctr}}=\Pi(\bm r_{\mathrm{ctr}})$. The range-azimuth pair is then defined as
\begin{align}
    r=\|\bar{\bm r}_{\mathrm{ctr}}\|_2,\qquad
    \theta=\mathrm{atan2}(y_c,x_c).
\end{align}
Given the observation tensor $\tilde{\mathcal H}_s$, the goal is to jointly infer
\begin{align}
    \{\hat c,\hat r,\hat\theta\}
    =
    f_{\Theta}(\tilde{\mathcal H}_s),
    \label{eq:joint_mapping_final}
\end{align}
where $f_{\Theta}(\cdot)$ is the learned mapping parameterized by $\Theta$.

This label associates localization with the geometric support of the object rather than with a single dominant scatterer.

To avoid angular discontinuity, the azimuth is encoded in sine-cosine form:
\begin{align}
    \bm a = [\sin\theta,\cos\theta]^{\mathsf T}.
\end{align}
The network predicts a normalized vector $\hat{\bm a}$, from which $\hat\theta$ is recovered as
\begin{align}
    \hat\theta = \mathrm{atan2}\big([\hat{\bm a}]_1,[\hat{\bm a}]_2\big).
\end{align}

Besides the individual range and azimuth errors, the final localization quality is characterized by the planar localization error
\begin{align}
    e_{{\rm pl},i}
    =
    \sqrt{
    \hat r_i^2+r_i^2
    -
    2\hat r_i r_i\cos(\Delta\theta_i)
    },
    \label{eq:planar_error_final}
\end{align}
where $\Delta\theta_i\triangleq\mathrm{wrap}(\hat\theta_i-\theta_i)\in[-\pi,\pi)$ denotes the wrapped azimuth error in radians. This metric measures the exact Euclidean localization error in the sensing plane and will later motivate the physics-guided regularization in Section~\ref{sec3}.

\section{Proposed Joint Learning Framework}\label{sec3}

\begin{figure*}[t]
\centering
\includegraphics[width=1\textwidth]{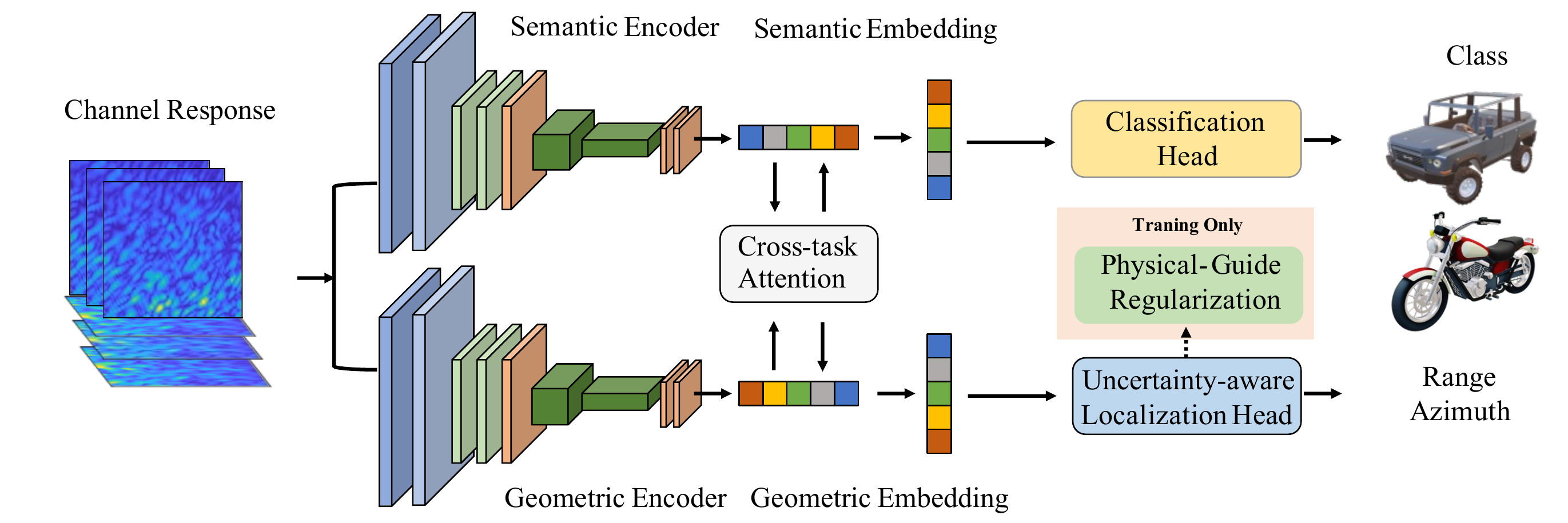}
\caption{Overall system framework of the proposed method.}
\label{fig:framework_overview}
\end{figure*}

The two tasks considered here are inferred from the same near-field tensor, but they do not rely on the same structures in that tensor. Classification depends more on distributed scattering signatures, while localization depends more on phase-aligned geometric variation across the array and the selected subcarriers. The framework in Fig.~\ref{fig:framework_overview} is built around this mismatch. Task-specific features are extracted first, cross-task interaction is introduced next, and localization is then regularized by losses that better match the final planar metric.

\subsection{Overall Architecture and Dual-Branch Feature Extraction}

The network input is the real-valued observation tensor defined in \eqref{eq:Hreal_final},
\begin{align}
    \tilde{\mathcal H}_s \in \mathbb{R}^{2\times N_r\times N_t\times K_s},
\end{align}
where the first dimension corresponds to the real and imaginary channels and the last dimension collects the selected subcarriers. Two task-specific branches are adopted to reduce feature conflict from the first stage of representation learning, in contrast to classical hard-parameter-sharing multi-task learning \cite{Caruana1997}, and in line with task-interaction designs such as Cross-Stitch and the Multi-Task Attention Network (MTAN) \cite{Misra2016,LiuMTAN2019}. Specifically, the network contains a semantic (classification-oriented) branch and a geometric (localization-oriented) branch.

Let $\mathcal E_{\mathrm{cls}}(\cdot)$ and $\mathcal E_{\mathrm{loc}}(\cdot)$ denote the semantic and geometric encoders, respectively. They take the same input tensor but use independent parameters:
\begin{align}
    \bm F_{\mathrm{cls}} = \mathcal E_{\mathrm{cls}}(\tilde{\mathcal H}_s), \qquad
    \bm F_{\mathrm{loc}} = \mathcal E_{\mathrm{loc}}(\tilde{\mathcal H}_s),
    \label{eq:dual_feature_maps}
\end{align}
where $\bm F_{\mathrm{cls}}, \bm F_{\mathrm{loc}} \in \mathbb{R}^{L\times d_f}$ are tokenized latent feature sequences, $L$ is the number of tokens after flattening the spatial-frequency feature maps, and $d_f$ is the token dimension. In the implementation used in Section~\ref{sec4}, each branch starts with three convolutional blocks, each composed of a 2-D convolution, batch normalization, and a rectified linear unit (ReLU), together with two intermediate max-pooling stages. The reduced spatial-frequency feature map is then flattened into a token sequence, so the token extractor is kept identical across branches and only the branch weights differ. The resulting sequence is refined by two transformer-style encoder blocks with four attention heads, layer normalization, Gaussian error linear unit (GELU) feed-forward layers of width $2d_f$, and learnable subcarrier embeddings. The semantic branch is intended to preserve morphology- and material-related scattering patterns, while the geometric branch targets phase-amplitude structures more directly tied to range and azimuth.

\subsection{Cross-Task Attention for Semantic-Geometric Interaction}

The two branches process the same near-field channel responses and thus retain complementary cues. Bidirectional cross-task attention is introduced at the token level following Transformer-style attention \cite{Vaswani2017}. From $\bm F_{\mathrm{cls}}$ and $\bm F_{\mathrm{loc}}$, the query, key, and value matrices are formed as
\begin{align}
    \bm Q_{\mathrm{cls}} &= \bm F_{\mathrm{cls}}\bm W_{Q}^{\mathrm{cls}}, &
    \bm K_{\mathrm{loc}} &= \bm F_{\mathrm{loc}}\bm W_{K}^{\mathrm{loc}}, &
    \bm V_{\mathrm{loc}} &= \bm F_{\mathrm{loc}}\bm W_{V}^{\mathrm{loc}}, \\
    \bm Q_{\mathrm{loc}} &= \bm F_{\mathrm{loc}}\bm W_{Q}^{\mathrm{loc}}, &
    \bm K_{\mathrm{cls}} &= \bm F_{\mathrm{cls}}\bm W_{K}^{\mathrm{cls}}, &
    \bm V_{\mathrm{cls}} &= \bm F_{\mathrm{cls}}\bm W_{V}^{\mathrm{cls}},
\end{align}
where $\bm W_{Q}^{(\cdot)}$, $\bm W_{K}^{(\cdot)}$, and $\bm W_{V}^{(\cdot)}$ are trainable projection matrices. The semantic tokens are refined by attending to the geometric tokens, while the geometric tokens are refined symmetrically:
\begin{align}
    \tilde{\bm F}_{\mathrm{cls}}
    &=
    \bm F_{\mathrm{cls}}
    +
    \mathrm{softmax}\!\left(
    \frac{\bm Q_{\mathrm{cls}}\bm K_{\mathrm{loc}}^{\mathsf T}}{\sqrt{d_a}}
    \right)\bm V_{\mathrm{loc}}, \\
    \tilde{\bm F}_{\mathrm{loc}}
    &=
    \bm F_{\mathrm{loc}}
    +
    \mathrm{softmax}\!\left(
    \frac{\bm Q_{\mathrm{loc}}\bm K_{\mathrm{cls}}^{\mathsf T}}{\sqrt{d_a}}
    \right)\bm V_{\mathrm{cls}},
    \label{eq:cross_task_attention}
\end{align}
where $d_a$ is the attention dimension. Residual connections are embedded in \eqref{eq:cross_task_attention}, followed by feed-forward refinement. In the implementation used in Section~\ref{sec4}, the residual contribution from the cross branch is further controlled by learnable mixing coefficients, initialized to a small value, so that the model is not forced into strong cross-task coupling at the start of training.

The attention weights in \eqref{eq:cross_task_attention} are computed from the current semantic and geometric features rather than fixed coefficients. The semantic branch thus receives location-sensitive cues from the geometric branch, while the geometric branch receives category-dependent scattering cues from the semantic branch.

The refined token sequences are then aggregated by global pooling and linear projection:
\begin{align}
    \bm h_{\mathrm{cls}} = \mathcal P_{\mathrm{cls}}(\tilde{\bm F}_{\mathrm{cls}}), \qquad
    \bm h_{\mathrm{loc}} = \mathcal P_{\mathrm{loc}}(\tilde{\bm F}_{\mathrm{loc}}),
    \label{eq:pooled_embeddings}
\end{align}
where $\bm h_{\mathrm{cls}}, \bm h_{\mathrm{loc}} \in \mathbb{R}^{d}$ are the task-level semantic and geometric embeddings. The semantic embedding is used for classification, while the geometric embedding is used for localization, uncertainty estimation, and the auxiliary regularization head.

\begin{remark}
If $\tilde{\bm F}_{\mathrm{cls}}=\bm F_{\mathrm{cls}}$ and $\tilde{\bm F}_{\mathrm{loc}}=\bm F_{\mathrm{loc}}$, the proposed architecture reduces to a task-decoupled dual-branch model. If the two branches are further collapsed into a shared encoder, the resulting structure approaches hard-parameter-sharing multi-task learning \cite{Caruana1997}. The proposed cross-task attention provides a controlled intermediate design between fully decoupled and fully shared representations.
\end{remark}

\subsection{Uncertainty-Aware Task Supervision}

Based on the task-level embeddings in \eqref{eq:pooled_embeddings}, shallow prediction heads are introduced for category prediction, range regression, and azimuth regression.

The classification head maps the semantic embedding to the class logits:
\begin{align}
    \bm o_{\mathrm{cls}} = \mathcal G_{\mathrm{cls}}(\bm h_{\mathrm{cls}}).
    \label{eq:cls_head}
\end{align}
The classification probability vector and the predicted label are then given by
\begin{align}
    \hat{\bm p}=\mathrm{softmax}(\bm o_{\mathrm{cls}}), \qquad
    \hat c=\arg\max_{c}[\hat{\bm p}]_c.
    \label{eq:cls_output}
\end{align}

The geometric embedding is used to predict the range and its log-variance:
\begin{align}
    \hat r = \mathcal G_r(\bm h_{\mathrm{loc}}), \qquad
    \hat s_r = \mathcal G_{u_r}(\bm h_{\mathrm{loc}}).
    \label{eq:range_head}
\end{align}

The azimuth branch outputs a sine-cosine representation together with its log-variance:
\begin{align}
    \hat{\bm a} = \mathcal G_{\theta}(\bm h_{\mathrm{loc}}), \qquad
    \hat s_\theta = \mathcal G_{u_\theta}(\bm h_{\mathrm{loc}}),
    \label{eq:azimuth_head}
\end{align}
where $\hat{\bm a}=[\widehat{\sin\theta},\widehat{\cos\theta}]^{\mathsf T}$ is normalized as
\begin{align}
    \hat{\bm a} \leftarrow \frac{\hat{\bm a}}{\|\hat{\bm a}\|_2},
    \label{eq:azimuth_norm}
\end{align}
and the azimuth estimate $\hat\theta$ is recovered from $\hat{\bm a}$ through the sine-cosine decoding introduced in Section~\ref{sec2}. During inference, only $\hat c$, $\hat r$, and $\hat\theta$ are retained.

The prediction heads are supervised by a task-oriented objective that combines category prediction with uncertainty-aware range-azimuth regression. The network estimates both the target parameters and their sample-dependent regression uncertainty \cite{Kendall2018}.

The task-supervision part is written as
\begin{align}
    \mathcal L_{\mathrm{task}}
    =
    \mathcal L_{\mathrm{cls}}
    +
    \lambda_r \mathcal L_r
    +
    \lambda_\theta \mathcal L_\theta,
    \label{eq:loss_task}
\end{align}
where $\lambda_r$ and $\lambda_\theta$ are weighting coefficients.

The classification term $\mathcal L_{\mathrm{cls}}$ is the standard cross-entropy loss,
\begin{align}
    \mathcal L_{\mathrm{cls}}
    =
    -\frac{1}{B}\sum_{i=1}^{B}
    \log
    \frac{\exp([\bm o_{\mathrm{cls},i}]_{c_i})}
    {\sum_{c=1}^{C}\exp([\bm o_{\mathrm{cls},i}]_{c})}.
\end{align}
For the localization task, heteroscedastic Gaussian negative log-likelihood (NLL) surrogates are adopted for range and normalized azimuth errors \cite{Kendall2018}. Let
$e_{r,i}=\hat r_i-r_i$ and $e_{\theta,i}=\bar{\Delta\theta}_i$, where
$\bar{\Delta\theta}_i=\Delta\theta_i/\pi$ and
$\Delta\theta_i=\mathrm{wrap}(\hat\theta_i-\theta_i)\in[-\pi,\pi)$. Then the range and azimuth terms $\mathcal L_r$ and $\mathcal L_\theta$ can be written compactly as
\begin{align}
    \mathcal L_\nu
    =
    \frac{1}{2B}\sum_{i=1}^{B}
    \left(
    e^{-\hat s_{\nu,i}} e_{\nu,i}^2 + \hat s_{\nu,i}
    \right),
    \quad \nu\in\{r,\theta\},
    \label{eq:loss_nll}
\end{align}
with the corresponding precision weights $w_{\nu,i}=e^{-\hat s_{\nu,i}}$. The additive log-variance term in \eqref{eq:loss_nll} prevents trivial inflation of the predicted uncertainty.

The uncertainty heads adapt the effective regression weights according to the residual scale. The NLL term is not an auxiliary output for post-processing; it directly reshapes the training signal seen by the localization branch.

\subsection{Physics-Guided Structured Regularization and Optimization}

Independent task losses do not explicitly capture the exact planar localization metric or the structural priors carried by the near-field response. A physics-guided regularization term is added to $\mathcal L_{\mathrm{task}}$, and the overall loss is written as
\begin{align}
    \mathcal L
    =
    \mathcal L_{\mathrm{task}}
    +
    \mathcal L_{\mathrm{phy}},
    \label{eq:loss_total_new}
\end{align}
with
\begin{align}
    \mathcal L_{\mathrm{phy}}
    &=
    \lambda_p \mathcal L_{\mathrm{pl}}
    +
    \lambda_{pk} \mathcal L_{\mathrm{peak}}
    +
    \lambda_c \mathcal L_{\mathrm{cp}},
    \label{eq:loss_split}
\end{align}
where $\lambda_p$, $\lambda_{pk}$, and $\lambda_c$ are weighting coefficients.

The three terms in $\mathcal L_{\mathrm{phy}}$ play different roles. $\mathcal L_{\mathrm{pl}}$ aligns training with the exact planar localization metric. $\mathcal L_{\mathrm{peak}}$ injects an observation-domain saliency prior extracted from the multi-frequency channel response. $\mathcal L_{\mathrm{cp}}$ regularizes the balance between radial and angular errors.

To align training with the final localization metric, a planar consistency term is introduced:
\begin{align}
    \mathcal L_{\mathrm{pl}}
    =
    \frac{1}{B}\sum_{i=1}^{B}
    \sqrt{
    \hat r_i^2+r_i^2
    -
    2\hat r_i r_i\cos(\Delta\theta_i)
    + \epsilon
    }.
    \label{eq:loss_planar_new}
\end{align}
Unlike independent squared penalties on range and azimuth, \eqref{eq:loss_planar_new} directly matches the exact Euclidean localization error in the sensing plane.

To support the peak-response regularization, an auxiliary peak head is introduced from the geometric embedding:
\begin{align}
    \hat{\bm u} = \mathcal G_{u}(\bm h_{\mathrm{loc}}),
    \label{eq:aux_head}
\end{align}
where $\hat{\bm u}\in\mathbb{R}^{2}$ is the predicted observation-domain peak coordinate used only during training.

To incorporate an observation-driven physical prior, an auxiliary peak-response constraint is constructed from the dominant saliency of the input. For the $i$-th sample, let
\begin{align}
    \bm M_i = \sum_{m=1}^{K_s}\big|\bm H_{k_m}^{(i)}\big|,
    \qquad
    \bm u_i^\star = \arg\max_{\bm u}\,[\bm M_i]_{\bm u},
    \label{eq:peak_reference}
\end{align}
where $\bm M_i$ is the aggregated magnitude map over the selected subcarriers and $\bm u_i^\star$ is the coordinate of its dominant peak. The corresponding peak-response prior is
\begin{align}
    \mathcal L_{\mathrm{peak}}
    =
    \frac{1}{B}\sum_{i=1}^{B}
    \|\hat{\bm u}_i-\bm u_i^\star\|_1,
\end{align}
where $\hat{\bm u}_i$ is the predicted peak coordinate generated by the auxiliary head in \eqref{eq:aux_head}. The summation over the selected subcarriers in \eqref{eq:peak_reference} suppresses isolated subcarrier fluctuations while preserving the dominant spatial saliency induced by the target. This prior is kept weak. It does not force the dominant magnitude peak to coincide with the final geometric center, but biases the localization branch toward physically active regions of the observation. To regularize the coupled nature of range and azimuth errors, a geometry-coupling term is also adopted:
\begin{align}
    \mathcal L_{\mathrm{cp}}
    =
    \frac{1}{B}\sum_{i=1}^{B}
    \Big|
    |\hat r_i-r_i|
    -
    |r_i\Delta\theta_i|
    \Big|.
\end{align}
Here, $|r_i\Delta\theta_i|$ corresponds to the tangential displacement measured along the sensing circle of radius $r_i$. Unlike $\mathcal L_{\mathrm{pl}}$, which measures the exact planar error, $\mathcal L_{\mathrm{cp}}$ is an auxiliary regularizer that discourages strong imbalance between radial and tangential localization discrepancies.

Training is end-to-end, while inference only requires a single forward pass through the learned network.

\section{Numerical Results}\label{sec4}

\subsection{Simulation Setup}

The simulation pipeline is based on electromagnetic data generation for sparse multi-frequency sensing. A cross-shaped near-field MIMO array with $N_t=N_r=64$ is considered at a carrier frequency of $4.9$\,GHz, and an orthogonal frequency-division multiplexing waveform with $K=16$ subcarriers is used for sensing. Realistic extended targets are generated from voxelized 3D vehicle models derived from ShapeNet \cite{ShapeNet}. The discretized VIE solver preserves distributed scattering and near-field wavefront effects beyond point-target abstractions. Background-suppressed target responses are considered, so homogeneous free space is used instead of a cluttered multipath scene. The dataset contains two classes, motorbike and car, with continuous range-azimuth labels. The projected target center is sampled over ranges from $5$ to $50$ m and azimuths from $-60^\circ$ to $60^\circ$. Metallic body parts and lossy tires are assigned different complex contrasts with random perturbations across samples, and each target is placed at a random heading angle. The resulting channel tensors exhibit substantial intra-class diversity, while the train/validation/test split remains mesh-disjoint: each 3D vehicle instance appears in only one split, so test samples are not new poses or material realizations of training meshes. The near-field channel responses are obtained by solving the discretized volume integral equation offline for each configuration.

Each sensing sample uses $K_s$ selected sensing subcarriers, and the default configuration adopts $K_s=16$. The network input is represented by the real-valued tensor $\tilde{\mathcal H}_s\in\mathbb{R}^{2\times N_r\times N_t\times K_s}$. Under the default setting, the input size is therefore $2\times 64\times 64\times 16$. The adopted dataset contains $3984$ training samples, $500$ validation samples, and $492$ test samples. The class counts are nearly balanced: the train/validation/test splits contain $1984/252/240$ motorbike samples and $2000/248/252$ car samples, respectively. Each sample is labeled by the target category together with the range and azimuth associated with the projected geometric center defined in Section~\ref{sec2}. The reduced-data experiments in Section~\ref{sec4d} are conducted by subsampling only the training split while keeping the validation and test sets unchanged.

For the benchmark tradeoff calibration used in the communication-oriented analysis below, we additionally instantiate \eqref{eq:Rerg} with an auxiliary shared-OFDM block of size $K_{\mathrm{tot}}=64$ and a reference full-band per-subcarrier SNR $\bar\gamma_c=15$~dB. This setting calibrates the horizontal rate axis and leaves the sensing dataset and learning models unchanged.

Table~\ref{tab:sim_setup} summarizes the main simulation and training parameters, including the Adam with decoupled weight decay (AdamW) optimizer.
\begin{table}[t]
\centering
\footnotesize
\renewcommand{\arraystretch}{1.05}
\caption{Main simulation and training parameters under the default setting.}
\label{tab:sim_setup}
\begin{tabular}{ll}
\hline
Carrier frequency & $4.9$ GHz \\
Transmit/receive antennas & $N_t=N_r=64$ \\
Array spacing & $\lambda_c/2$ \\
Default input tensor size & $2\times 64\times 64\times 16$ \\
Default selected subcarriers & $K_s=16$ \\
Rate-proxy benchmark & $K_{\mathrm{tot}}=64$, $\bar\gamma_c=15$ dB \\
Target models & ShapeNet-derived vehicles \\
Number of classes & $C=2$ (motorbike/car) \\
Intra-class randomization & heading, material contrast, position \\
Training/validation/test samples & $3984/500/492$ \\
Labels & class, range, azimuth \\
Optimizer & AdamW \\
Initial learning rate & $2\times 10^{-4}$ \\
Training epochs & $200$ \\
Seeds & $\{42,43,44\}$ \\
\hline
\end{tabular}
\end{table}

The proposed model is trained in a supervised manner using AdamW \cite{LoshchilovHutter2019} with an initial learning rate of $2\times 10^{-4}$ and cosine-annealing learning-rate decay \cite{LoshchilovHutter2017}. Unless otherwise specified, the default training length is $200$ epochs, the loss weights are $(\lambda_r,\lambda_\theta,\lambda_p,\lambda_{pk},\lambda_c)=(1.0,1.8,0.2,0.05,0.1)$, and the reported results are averaged over three seeds $\{42,43,44\}$. Mild sample-wise amplitude perturbation is applied only to the training split. Model selection is performed on the validation set using the same criterion throughout the experiments, and all tables and figures in this section report final performance on the test split unless explicitly stated otherwise. The reported metrics are classification accuracy, range mean absolute error (MAE), azimuth MAE, planar MAE, and the success rate within 1 m, denoted by $\mathrm{Succ}@1\mathrm{m}$.

\subsection{Comparison With Baselines}
\begin{figure}[!t]
\centering
\includegraphics[width=0.9\columnwidth]{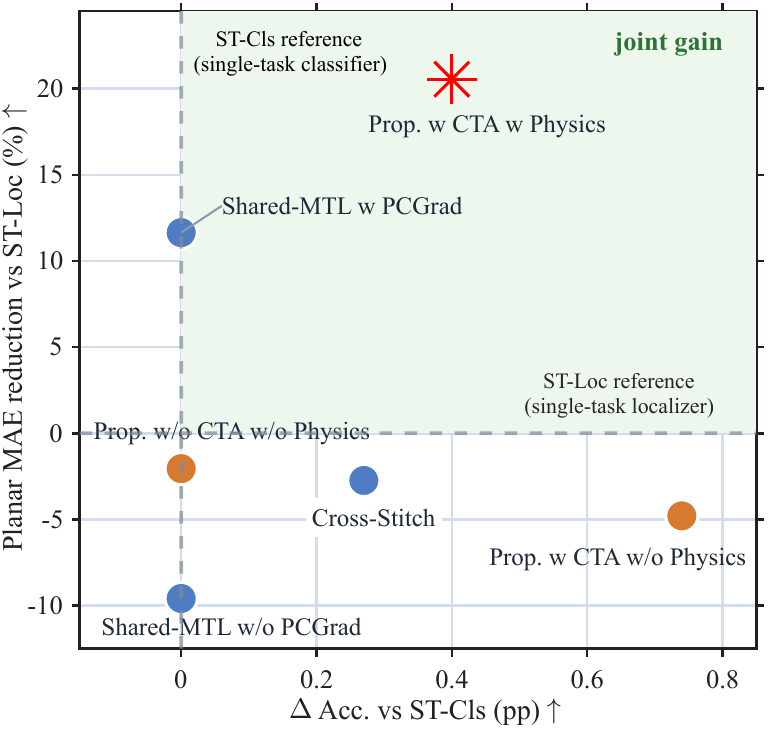}
\caption{Joint-learning gain map relative to the single-task references.}
\label{fig:overall_summary}
\end{figure}

Representative baselines under the default setting are compared next, and Fig.~\ref{fig:overall_summary} visualizes the gain map. The compared methods fall into four groups: two deterministic model-based references, the far-field fast Fourier transform (FFT) periodogram, denoted as Traditional FFT (2D-Periodogram), rooted in classical spectral estimation \cite{Kay1993}, and a near-field point-target matched filter, which follows spherical-wave point-target localization ideas in near-field array processing \cite{Bekkerman2006,Wang2025NFLS}; shared-encoder multi-task baselines, Shared-MTL w/o PCGrad \cite{Caruana1997} and Shared-MTL w PCGrad \cite{Caruana1997,YuPCGrad2020}, where the latter applies the standard PCGrad update \cite{YuPCGrad2020} to the shared-encoder backbone; a soft-sharing multi-task baseline, Cross-Stitch \cite{Misra2016}; and two single-task references for classification and localization. The reduced proposed variants, denoted as Proposed w/o CTA w/o physics and Proposed w CTA w/o physics, are reported separately to show how the gain evolves inside the same backbone, while Proposed w CTA w physics denotes the complete framework. The external learning baselines are kept under their standard deterministic objectives, whereas the proposed-family comparison is used to separate the effects of controlled task interaction and structured supervision within the same backbone. Since the two classical model-based references do not involve training, they are reported as single test-set means rather than seed-averaged mean$\pm$standard-deviation values.

\begin{table*}[t]
\caption{Comparison with baselines under the default setting.}
\label{tab:overall_comparison}
\centering
\footnotesize
\renewcommand{\arraystretch}{1.05}
\setlength{\tabcolsep}{4.0pt}
\begin{tabular}{>{\raggedright\arraybackslash}p{4.80cm}ccccc}
\hline
Method & Acc. (\%) & Range MAE (m) & Azimuth MAE ($^\circ$) & Planar MAE (m) & Succ@1m (\%) \\
\hline
\multicolumn{6}{l}{\textit{Classical model-based references}} \\
Traditional FFT (2D-Periodogram) & -- & 16.4085 & 1.4270 & 16.4389 & 2.64 \\
Near-field point-target matched filter & -- & 8.8971 & 1.6732 & 8.9847 & 6.30 \\
\hline
\multicolumn{6}{l}{\textit{Joint-learning baselines}} \\
Shared-MTL w/o PCGrad \cite{Caruana1997} & 98.92$\pm$0.35 & 1.0535$\pm$0.0160 & 2.0537$\pm$0.0939 & 1.6033$\pm$0.0560 & 39.23$\pm$1.68 \\
Shared-MTL w PCGrad \cite{Caruana1997,YuPCGrad2020} & 98.92$\pm$0.25 & 1.0650$\pm$0.0165 & 1.2158$\pm$0.0257 & 1.2912$\pm$0.0205 & 50.27$\pm$0.82 \\
Cross-Stitch \cite{Misra2016} & 99.19$\pm$0.33 & 1.0580$\pm$0.0282 & 1.8163$\pm$0.0459 & 1.4969$\pm$0.0106 & 43.02$\pm$0.42 \\
\hline
\multicolumn{6}{l}{\textit{Single-task references}} \\
Single-task classification & 98.92$\pm$0.42 & -- & -- & -- & -- \\
Single-task localization & -- & 1.0528$\pm$0.0184 & 1.7529$\pm$0.0530 & 1.4630$\pm$0.0324 & 43.29$\pm$0.17 \\
\hline
\multicolumn{6}{l}{\textit{Proposed variants}} \\
Proposed w/o CTA w/o physics & 98.92$\pm$0.10 & \textbf{1.0427$\pm$0.0343} & 1.8439$\pm$0.0748 & 1.4884$\pm$0.0568 & 39.70$\pm$2.46 \\
Proposed w CTA w/o physics & \textbf{99.66$\pm$0.25} & 1.0691$\pm$0.0091 & 1.8566$\pm$0.1238 & 1.5344$\pm$0.0309 & 42.28$\pm$1.45 \\
Proposed w CTA w physics & 99.32$\pm$0.10 & 1.0787$\pm$0.0137 & \textbf{0.7299$\pm$0.0386} & \textbf{1.1641$\pm$0.0174} & \textbf{57.79$\pm$1.07} \\
\hline
\end{tabular}
\end{table*}

Among all compared methods, the proposed framework achieves the lowest planar MAE and the highest $\mathrm{Succ}@1\mathrm{m}$. The two classical model-based baselines already reveal the importance of matching the sensing geometry to the operating regime. Specifically, Table~\ref{tab:overall_comparison} shows that moving from the far-field 2D periodogram to the near-field point-target matched filter reduces the planar MAE substantially, yet the remaining gap to the proposed method is still large because realistic extended targets cannot be represented well by a single point scatterer. Among the learning-based methods, Proposed w CTA w/o physics attains the highest classification accuracy, which indicates that cross-task attention effectively transfers geometric cues to semantic discrimination. Its localization metrics remain clearly worse than those of Shared-MTL w PCGrad, suggesting that CTA alone pushes the joint model toward a classification-favorable operating point. Once uncertainty-aware regression and the physics-guided terms are added, the same backbone achieves the best azimuth MAE, the best planar MAE, and the highest $\mathrm{Succ}@1\mathrm{m}$ while maintaining highly competitive classification accuracy. The full model combines CTA-driven classification gains with a structured objective that restores a more balanced joint classifier-localizer.

The gain map further visualizes the multi-task benefit relative to the two single-task references: the horizontal axis measures the classification-accuracy gain over Single-task classification, while the vertical axis measures the relative planar-MAE reduction over Single-task localization. The upper-right region corresponds to a desirable synergistic regime in which classification accuracy is improved over the single-task classifier while planar localization error is simultaneously reduced relative to the single-task localizer. In the gain map, Proposed w CTA w/o physics mainly shifts rightward, reflecting the classification gain brought by cross-task interaction. The full proposed model then moves further upward after the uncertainty-aware and physics-guided objective is introduced. CTA primarily helps classification, while the structured objective pulls the joint model into the strongest balanced operating region.

\subsection{Performance Under Limited Frequency Resources}

\begin{figure*}[t]
\centering
\includegraphics[width=\textwidth]{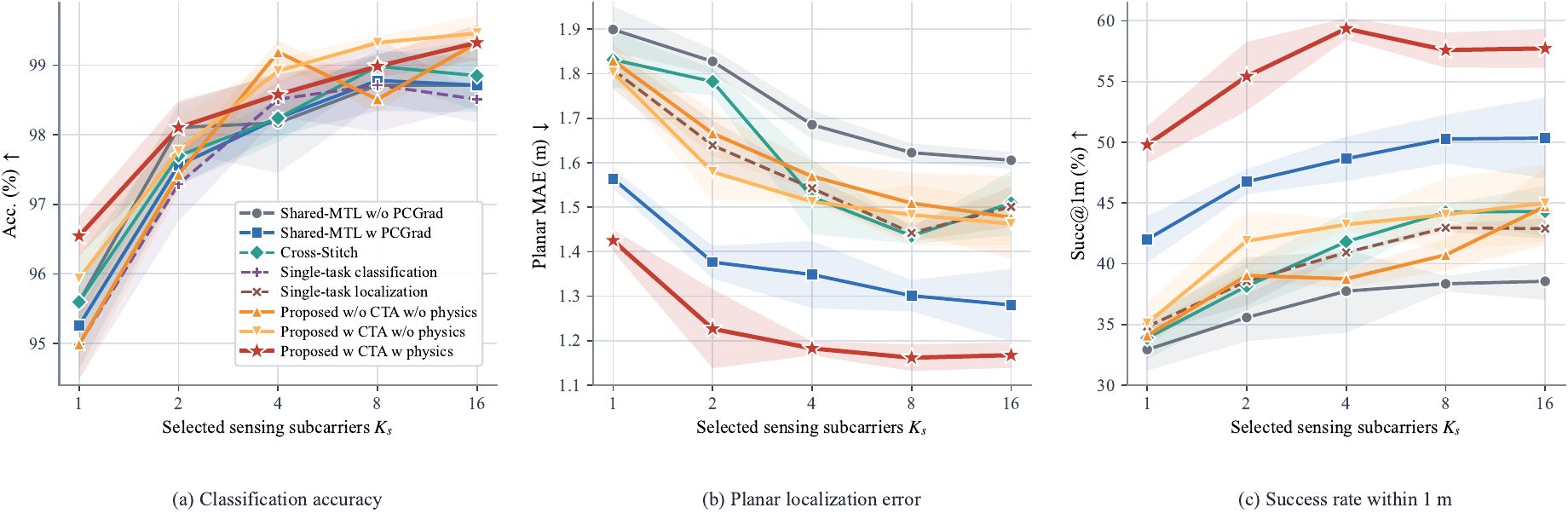}
\caption{Performance versus the number of selected subcarriers $K_s$.}
\label{fig:freq_resource}
\end{figure*}

The next comparison examines the behavior of the complete proposed model and the above baselines when the number of sensing subcarriers is reduced. All methods are evaluated at $K_s\in\{1,2,4,8,16\}$, while the total OFDM resource remains fixed. This comparison reveals whether accurate joint inference can still be maintained when the sensing-side frequency budget is limited, and it also reflects the sensing-communication resource trade-off because increasing the sensing allocation reduces the residual communication resource. To keep the visualization readable while still covering the main trends, Fig.~\ref{fig:freq_resource} reports three representative metrics: classification accuracy, planar MAE, and $\mathrm{Succ}@1\mathrm{m}$. All learning-based baselines and the two single-task references with complete frequency-sweep experiments are included. The classical model-based references are omitted because only their default $K_s=16$ evaluations are available.

The frequency sweep reveals a consistent ordering. In Fig.~\ref{fig:freq_resource}(b) and Fig.~\ref{fig:freq_resource}(c), Proposed w CTA w physics forms the dominant localization frontier across all tested $K_s$, with especially visible gains when the sensing-side frequency budget is highly constrained. For example, at $K_s=4$ it already reaches a planar MAE of $1.1826$~m and a $\mathrm{Succ}@1\mathrm{m}$ of $59.35\%$, while the strongest non-proposed baseline Shared-MTL w PCGrad remains at $1.3485$~m and $48.64\%$, respectively. Figure~\ref{fig:freq_resource}(a) further shows that the classification accuracy becomes saturated for several methods once $K_s\ge 4$, so the main practical difference under limited frequency resources lies in localization fidelity rather than in recognition alone. These results show that the proposed task-aware architecture and structured objective are especially beneficial when the frequency resources allocated to sensing are limited and motivate the benchmark-rate view studied next.

\subsection{Sensing-Communication Tradeoff}

The limited-frequency study above can also be expressed on a communication-oriented axis by mapping each sensing allocation $K_s$ to the ergodic-rate measure $R_{\mathrm{erg}}(K_s)$ in \eqref{eq:Rerg}. This subsection instantiates the bandwidth-allocation viewpoint in \eqref{eq:qos_rate_alloc}: smaller sensing allocations that still satisfy a target sensing-QoS region correspond to more communication-favorable operating points. Here we use the auxiliary shared-OFDM setting described earlier, namely $K_{\mathrm{tot}}=64$ total subcarriers and reference full-band per-subcarrier SNR $\bar\gamma_c=15$~dB. The resulting rate axis lets the sensing-bandwidth sweep be read directly in communication-resource terms.

\begin{figure}[t]
\centering
\includegraphics[width=\columnwidth]{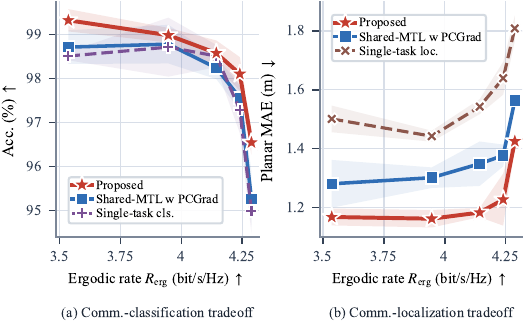}
\caption{Sensing-communication tradeoff under the ergodic-rate benchmark.}
\label{fig:comm_tradeoff}
\end{figure}

Fig~\ref{fig:comm_tradeoff} makes this tradeoff explicit. The left panel depicts the rate/classification relation and the right panel depicts the rate/localization relation, with the shaded bands indicating seed-to-seed variation. In Fig.~\ref{fig:comm_tradeoff}(a), the proposed model remains on the favorable accuracy frontier across the considered rates and consistently stays above the strongest joint-learning baseline Shared-MTL w PCGrad. In Fig.~\ref{fig:comm_tradeoff}(b), the separation is even clearer for localization: at the same $R_{\mathrm{erg}}$, the proposed model achieves the lowest planar MAE throughout the sweep. For instance, at $K_s=4$, which corresponds to $R_{\mathrm{erg}}\approx 4.14$~bit/s/Hz in this setting, the proposed method already attains $98.58\%$ classification accuracy together with a planar MAE of $1.1826$~m, whereas Shared-MTL w PCGrad remains at $98.24\%$ and $1.3485$~m. From the perspective of \eqref{eq:qos_rate_alloc}, this means that the proposed model reaches the same sensing-quality region at a smaller sensing allocation and hence at a more communication-favorable point on the rate axis. Even at $K_s=2$ ($R_{\mathrm{erg}}\approx 4.24$~bit/s/Hz), the proposed framework still maintains a clear localization advantage.

To make the communication-efficiency implication explicit, Table~\ref{tab:qos_comm_efficiency} reports the minimum sensing bandwidth $K_s^\star$ required by representative joint-learning methods to satisfy two sensing-QoS targets, together with the associated rate. Smaller $K_s^\star$ means fewer reserved sensing tones and a higher rate under \eqref{eq:Rerg}. Here, QoS-I is defined by $A(K_s)\ge 98.0\%$ and $E_{\mathrm p}(K_s)\le 1.50$~m, while QoS-II is defined by $A(K_s)\ge 98.5\%$ and $E_{\mathrm p}(K_s)\le 1.30$~m. For compactness, Table~\ref{tab:qos_comm_efficiency} abbreviates the reduced proposed variants as Proposed w/o CTA and Proposed w CTA, corresponding to Proposed w/o CTA w/o physics and Proposed w CTA w/o physics, respectively. The proposed full model reaches QoS-I with only $K_s^\star=2$, corresponding to $R_{\mathrm{erg}}\approx 4.24$~bit/s/Hz in this setting, while Shared-MTL w PCGrad requires $K_s^\star=4$ and Cross-Stitch requires $K_s^\star=8$. Under the stricter QoS-II target, the proposed full model remains feasible at $K_s^\star=4$, while Shared-MTL w PCGrad needs the full $K_s=16$ sweep endpoint. Relative to the competing joint-learning methods, the proposed model reaches the same target sensing-QoS region at a more communication-favorable operating point. Entries marked ``Not met'' indicate that the corresponding target is not achieved within the tested sweep $K_s\in\{1,2,4,8,16\}$.

\begin{table}[t]
\caption{Communication-efficient operating points under the two sensing-QoS targets defined in the text.}
\label{tab:qos_comm_efficiency}
\centering
\footnotesize
\renewcommand{\arraystretch}{1.04}
\setlength{\tabcolsep}{4.2pt}
\begin{tabular*}{\columnwidth}{@{\extracolsep{\fill}}lcccc}
\hline
Method & \multicolumn{2}{c}{QoS-I} & \multicolumn{2}{c}{QoS-II} \\
 & $K_s^\star$ & Bench. rate & $K_s^\star$ & Bench. rate \\
\hline
Shared-MTL w PCGrad & 4 & 4.1436 & 16 & 3.5361 \\
Cross-Stitch & 8 & 3.9467 & \multicolumn{2}{c}{Not met} \\
Proposed w/o CTA & 16 & 3.5361 & \multicolumn{2}{c}{Not met} \\
Proposed w CTA & 8 & 3.9467 & \multicolumn{2}{c}{Not met} \\
Proposed full & \textbf{2} & \textbf{4.2400} & \textbf{4} & \textbf{4.1436} \\
\hline
\end{tabular*}
\end{table}

Within this benchmark calibration, reserving only a few sensing tones, e.g., $K_s\le 4$ within the $64$-tone shared OFDM block, already preserves strong joint classification and localization performance while keeping the associated benchmark rate above $4$~bit/s/Hz. The next question is which architectural and objective components produce this operating-point advantage.

\subsection{Ablation Studies}

The ablation study analyzes the proposed design from two complementary perspectives. The architecture ablation compares representative joint-learning backbones under the basic deterministic objective. The objective ablation then fixes the dual-branch cross-task backbone and studies how the heteroscedastic NLL term and the structured physics-guided constraints reshape the final performance. To keep the presentation compact, both comparisons are summarized in two blocks.

\begin{table*}[t]
\caption{Ablation studies under the default setting.}
\label{tab:ablation_studies}
\centering
\footnotesize
\renewcommand{\arraystretch}{1.04}
\begin{tabular}{>{\raggedright\arraybackslash}p{4.0cm}ccccc}
\hline
Variant & Acc. (\%) & Range MAE (m) & Azimuth MAE ($^\circ$) & Planar MAE (m) & Succ@1m (\%) \\
\hline
\multicolumn{6}{l}{\textit{(a) Architecture variants with the basic deterministic objective}} \\
\hline
Shared-MTL w/o PCGrad \cite{Caruana1997} & 98.92$\pm$0.35 & 1.0535$\pm$0.0160 & 2.0537$\pm$0.0939 & 1.6033$\pm$0.0560 & 39.23$\pm$1.68 \\
Shared-MTL w PCGrad \cite{Caruana1997,YuPCGrad2020} & 98.92$\pm$0.25 & 1.0650$\pm$0.0165 & \textbf{1.2158$\pm$0.0257} & \textbf{1.2912$\pm$0.0205} & \textbf{50.27$\pm$0.82} \\
Cross-Stitch \cite{Misra2016} & 99.19$\pm$0.33 & 1.0580$\pm$0.0282 & 1.8163$\pm$0.0459 & 1.4969$\pm$0.0106 & 43.02$\pm$0.42 \\
Proposed w/o CTA w/o physics & 98.92$\pm$0.10 & \textbf{1.0427$\pm$0.0343} & 1.8439$\pm$0.0748 & 1.4884$\pm$0.0568 & 39.70$\pm$2.46 \\
Proposed w CTA w/o physics & \textbf{99.66$\pm$0.25} & 1.0691$\pm$0.0091 & 1.8566$\pm$0.1238 & 1.5344$\pm$0.0309 & 42.28$\pm$1.45 \\
\hline
\multicolumn{6}{l}{\textit{(b) Objective variants on the dual-branch cross-task backbone}} \\
\hline
Proposed w CTA w/o physics & \textbf{99.66$\pm$0.25} & \textbf{1.0691$\pm$0.0091} & 1.8566$\pm$0.1238 & 1.5344$\pm$0.0309 & 42.28$\pm$1.45 \\
Proposed w CTA w NLL only & 99.25$\pm$0.10 & 1.0992$\pm$0.0173 & \textbf{0.7281$\pm$0.0065} & 1.1810$\pm$0.0160 & \textbf{57.79$\pm$0.91} \\
w/o planar consistency & 98.78$\pm$0.33 & 1.1199$\pm$0.0459 & 0.8178$\pm$0.0304 & 1.2272$\pm$0.0452 & 55.89$\pm$0.88 \\
w/o peak regularization & 98.98$\pm$0.17 & 1.1177$\pm$0.0456 & 0.7428$\pm$0.0230 & 1.2006$\pm$0.0461 & 55.62$\pm$1.77 \\
w/o coupling constraint & 99.05$\pm$0.67 & 1.0957$\pm$0.0325 & 0.7516$\pm$0.0393 & 1.1821$\pm$0.0376 & 57.72$\pm$3.03 \\
Proposed w CTA w physics & 99.32$\pm$0.10 & 1.0787$\pm$0.0137 & 0.7299$\pm$0.0386 & \textbf{1.1641$\pm$0.0174} & \textbf{57.79$\pm$1.07} \\
\hline
\end{tabular}
\end{table*}

\subsubsection{Architecture Ablation}

The upper block evaluates whether task interaction alone is sufficient under the basic deterministic loss. Proposed w CTA w/o physics attains the highest classification accuracy, which suggests that controlled cross-task interaction allows localization-related geometric cues to assist semantic discrimination. Its localization metrics, however, remain inferior to those of Shared-MTL w PCGrad, indicating that interaction alone does not fully preserve the geometry-sensitive structure required for regression. The results therefore support task-specific encoding first and motivate additional structured supervision on the same CTA backbone.

\subsubsection{Objective Ablation}

The lower block of Table~\ref{tab:ablation_studies} then fixes the dual-branch cross-task backbone and varies only the training objective. Moving from Proposed w CTA w/o physics to Proposed w CTA w NLL only reduces the planar MAE from $1.5344$~m to $1.1810$~m, improves $\mathrm{Succ}@1\mathrm{m}$ from $42.28\%$ to $57.79\%$, and sharply lowers the azimuth MAE. This indicates that uncertainty-aware supervision is important once cross-task interaction is enabled. Adding the structured physics-guided terms then further reduces the planar MAE to $1.1641$~m, keeps $\mathrm{Succ}@1\mathrm{m}$ at $57.79\%$, and slightly recovers the classification accuracy from $99.25\%$ to $99.32\%$, although the NLL-only variant remains marginally better on azimuth MAE. Taken together, the ablation suggests that CTA mainly helps the classification side, whereas the additional structured terms mainly stabilize localization on the same backbone.

\subsection{Error Distribution Analysis}

The ablation results explain why the average metrics improve, but they do not show where the errors occur spatially. We next compare the spatial distribution of the planar localization error for the main overall-comparison methods that produce localization outputs. The set includes Traditional FFT, the near-field point-target matched filter, Single-task localization, Shared-MTL w PCGrad, Cross-Stitch, and Proposed w CTA w physics. Since Single-task classification does not output coordinates, it is not included in this figure. For visualization, the three-seed learning-based results are aggregated and then binned and interpolated over the range-azimuth plane to form continuous sector maps, while the deterministic model-based references are plotted from their single full-test evaluations. This analysis complements the mean-error and success-rate results by showing where large errors concentrate in the sensing sector.

Because classification errors are much sparser than localization errors, directly plotting a spatial misclassification-rate map would be statistically unstable. We instead compare the predictive classification entropy of the main classification-output methods, Single-task classification, Shared-MTL w/o PCGrad, Shared-MTL w PCGrad, Cross-Stitch, Proposed w CTA w/o physics, and Proposed w CTA w physics, while directly overlaying the misclassified samples. For visualization, the three-seed test-set outputs are aggregated for each method. In these predictive-entropy sector maps, the shared-backbone baselines exhibit broader elevated-entropy regions near the outer boundary and in several off-boresight sectors. Cross-Stitch and Proposed w CTA w/o physics already produce more concentrated uncertainty regions, whereas the full proposed model yields the fewest misclassified samples and keeps the high-entropy zones more spatially localized around the difficult regions. The remaining high-entropy hot spot is mainly concentrated in a small close-range region. One plausible explanation is that, under the adopted sensing geometry, the observed target extent can be incomplete there, which weakens the semantic cues used to distinguish motorbike from car.

\begin{figure*}[!t]
\centering
\includegraphics[width=0.985\textwidth]{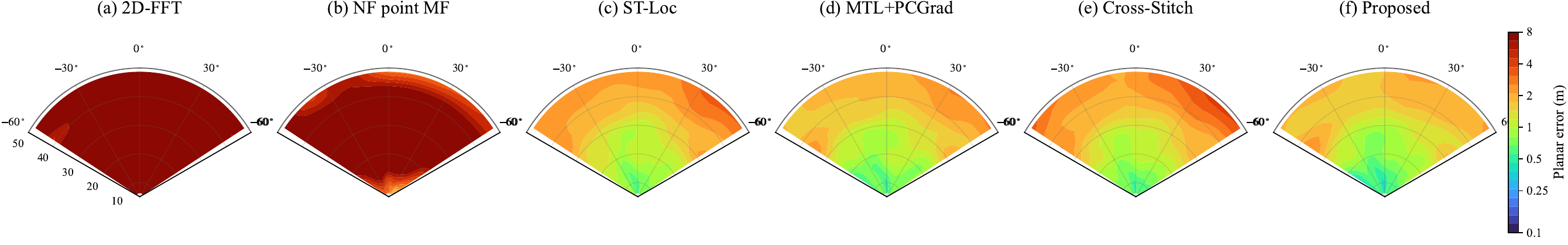}
\caption{Sector-wise planar-error maps for the main localization-output methods.}
\label{fig:error_heatmap}
\includegraphics[width=0.985\textwidth]{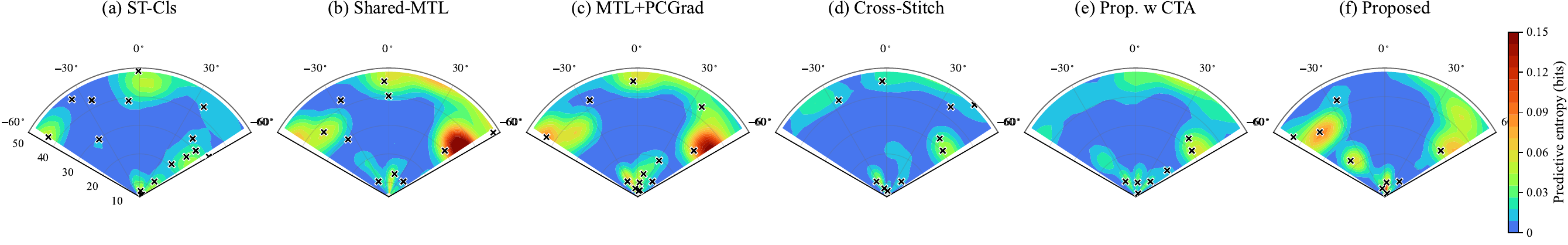}
\caption{Predictive-entropy sector maps for the main classification-output methods on the aggregated three-seed test split. Black crosses denote misclassified samples.}
\label{fig:cls_entropy_map}
\vspace{-0.8em}
\end{figure*}

The sector-wise error and entropy patterns further clarify where the performance gaps arise. Figure~\ref{fig:error_heatmap} shows that the two classical model-based references retain broad high-error regions over large portions of the sensing sector, especially at longer ranges, which is consistent with their much larger planar MAE in Table~\ref{tab:overall_comparison}. The learned baselines reduce these hot regions considerably, with Shared-MTL w PCGrad and Cross-Stitch already exhibiting much more controlled spatial behavior than the classical approaches. The proposed full model further yields the most uniformly low-error sector among the compared methods, with the clearest improvement appearing in the long-range and large-azimuth regions. This suggests that the physics-guided constraints are particularly effective at suppressing the geometric regression errors that tend to accumulate in those more challenging parts of the sensing sector. The complementary predictive-entropy view in Fig.~\ref{fig:cls_entropy_map} is consistent with this picture and keeps the uncertainty concentrated around the remaining difficult regions.

\subsection{Robustness to Reduced Training Data}\label{sec4d}

Reduced-data robustness is examined next. The training-data fraction is varied over multiple levels while keeping the validation and test splits unchanged. For readability, three representative metrics are reported over the practically relevant range from $25\%$ to $100\%$.

\begin{figure*}[t]
\centering
\includegraphics[width=\textwidth]{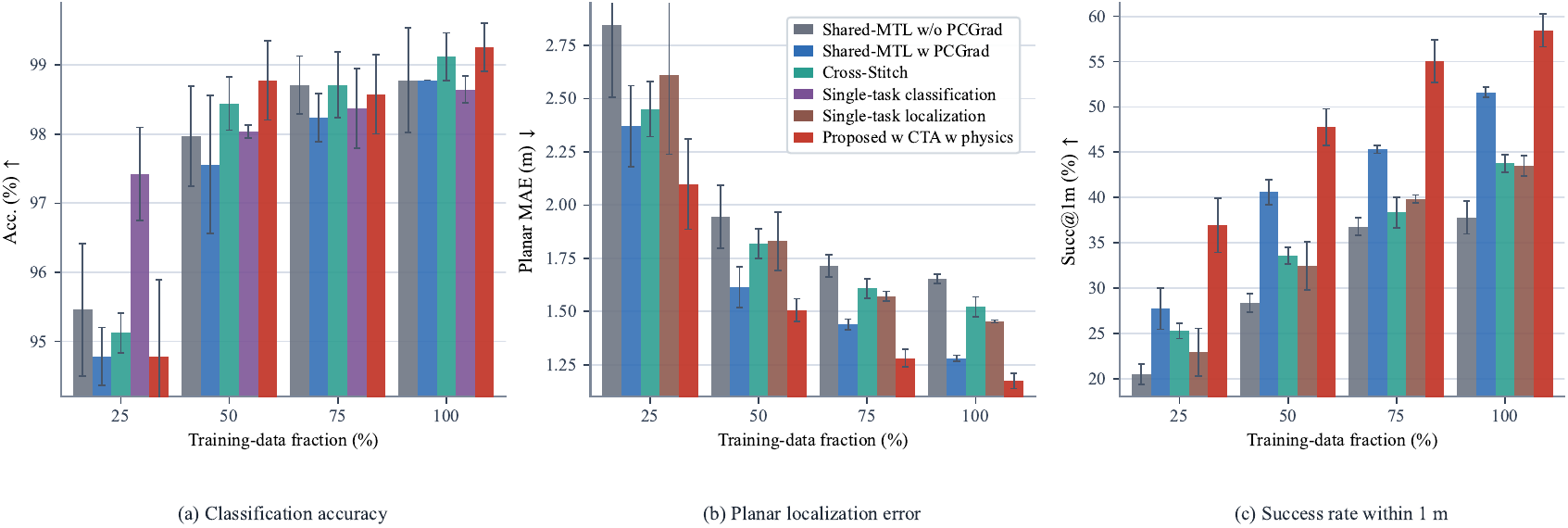}
\caption{Representative performance versus the training-data fraction under reduced-data conditions.}
\label{fig:data_fraction}
\end{figure*}

All methods improve as the training-data fraction increases, while the proposed family remains particularly robust in the localization-oriented metrics. As shown in Fig.~\ref{fig:data_fraction}, already at the lowest displayed fraction of $25\%$, the full model provides the clearest advantage in planar localization and $\mathrm{Succ}@1\mathrm{m}$, which suggests that the structured objective acts as a useful regularizer when supervision is limited. As the training set grows, the classification accuracies of several methods become more tightly clustered, while the full proposed model keeps the best localization accuracy across all fractions. The advantage is especially meaningful in the low-data regime, where accurate localization is harder to maintain.

\subsection{Sensitivity to Imperfect Input Responses}

\begin{figure}[t]
\centering
\includegraphics[width=\columnwidth]{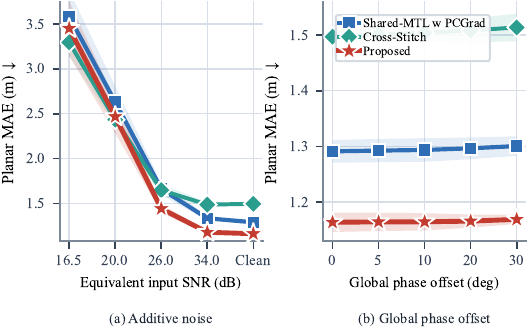}
\caption{Localization sensitivity under additive noise and global phase offset.}
\label{fig:input_robustness}
\end{figure}

The reduced-data experiment above addresses limited supervision. The final experiment studies robustness to imperfect test-time inputs. Starting from the default checkpoints, we perturb the input tensor only during inference and compare three representative joint models: Shared-MTL w PCGrad, Cross-Stitch, and Proposed w CTA w physics. We focus on two representative perturbation types: additive complex noise with normalized scale $\sigma/\mathrm{RMS}(H)\in\{0,0.02,0.05,0.10,0.15\}$, where RMS denotes the root-mean-square magnitude of the clean input, and global phase offsets in $\{0^\circ,5^\circ,10^\circ,20^\circ,30^\circ\}$. In Fig.~\ref{fig:input_robustness}(a), the noise strength is reported through the equivalent input signal-to-noise ratio axis $\mathrm{SNR}_{\mathrm{eq}}=-20\log_{10}(\sigma/\mathrm{RMS}(H))$, with the zero-noise point shown as the clean reference. For compactness, Fig.~\ref{fig:input_robustness} reports the seed-averaged planar MAE, which is the most sensitive localization metric in this perturbation study.

Figure~\ref{fig:input_robustness} shows that the two perturbation types have clearly different effects. Under global phase offsets, the proposed model is largely invariant and keeps the lowest planar MAE throughout the full sweep, which indicates good tolerance to moderate phase-calibration errors. Additive noise is more challenging for all methods, yet the proposed model preserves the best clean-condition localization and remains strongest over the moderate-SNR regime, while Cross-Stitch becomes slightly better only at the lowest tested SNR point. The uncertainty-aware and physics-guided objective improves nominal localization accuracy and maintains competitive stability under imperfect input responses, especially when the perturbation level is not extreme. Stronger clutter variation, synchronization mismatch, and hardware effects are left for future study.

\section{Conclusion} \label{sec5}
In this paper, receiver-side joint target classification and range-azimuth localization were studied for realistic extended targets from channel responses of a near-field ISAC system under a shared sensing-communication resource budget. A physics-grounded inference formulation was established by combining electromagnetic scattering modeling with a multi-frequency observation representation. On top of this formulation, a dual-branch framework with controlled cross-task interaction was developed. The numerical results point to a consistent picture: cross-task attention mainly helps classification, while the uncertainty-aware and physics-guided objective makes the same backbone reliable for localization. Under the adopted shared-OFDM rate interpretation, the proposed model reaches the same sensing-QoS region with a smaller sensing-bandwidth allocation. The results are obtained from full-wave simulation with background-suppressed channel responses and two target classes. Extensions to hardware measurements, cluttered and multipath-rich environments, richer target taxonomies, and tighter closed-loop communication-sensing integration are natural next steps.

\balance

\end{document}